\documentclass[conference]{IEEEtran}
\usepackage{fancyhdr} 

\usepackage{graphicx}
\usepackage{subcaption}
\usepackage{multirow}
\usepackage{booktabs}
\usepackage{pgfplots}
\pgfplotsset{compat=1.18}
\usepackage{xcolor}
\usepackage[hidelinks]{hyperref}

\graphicspath{{figures/}{diagrams/}}

\definecolor{darkgreen}{rgb}{0.0, 0.5, 0.0}

\makeatletter
\newcommand{\linebreakand}{%
  \end{@IEEEauthorhalign}\hfill\mbox{}\par
  \mbox{}\hfill\begin{@IEEEauthorhalign}
}
\makeatother

\begin{document}
\bstctlcite{IEEEexample:BSTcontrol}       

\title{Towards Efficient HPC Systems for Agents: Challenges and Opportunities}

\author{
    \IEEEauthorblockN{Yunjia Zheng}
    \IEEEauthorblockA{
    Harvard University\\
    \href{mailto:yunjia_zheng@g.harvard.edu}{yunjia\_zheng@g.harvard.edu}
    }
    \and
    \IEEEauthorblockN{Bintang Dwi Marthen}
    \IEEEauthorblockA{
    Harvard University\\
    \href{mailto:mmarthen@g.harvard.edu}{mmarthen@g.harvard.edu}
    }
    \and
    \IEEEauthorblockN{Zachary Pan}
    \IEEEauthorblockA{
    Harvard University\\
    \href{mailto:zapan@college.harvard.edu}{zapan@college.harvard.edu}
    }
            \linebreakand
    \and
    \IEEEauthorblockN{Minghao Li}
    \IEEEauthorblockA{
    Harvard University\\
    \href{mailto:minghaoli@g.harvard.edu}{minghaoli@g.harvard.edu}
    }

    \and
    \IEEEauthorblockN{Raminder Singh}
    \IEEEauthorblockA{
    Harvard FAS Research Computing\\
    \href{mailto:r_singh@g.harvard.edu}{r\_singh@g.harvard.edu}
    }
    \and
    \IEEEauthorblockN{Manasvita Joshi}
    \IEEEauthorblockA{
    Harvard FAS Research Computing\\
    \href{mailto:manasvitajoshi@g.harvard.edu}{manasvitajoshi@g.harvard.edu}
    }
            \linebreakand
    \and
    \IEEEauthorblockN{Minlan Yu}
    \IEEEauthorblockA{
    Harvard University\\
    \href{mailto:minlanyu@g.harvard.edu}{minlanyu@g.harvard.edu}
    }
    \and
    \IEEEauthorblockN{Juncheng Yang}
    \IEEEauthorblockA{
    Harvard University\\
    \href{mailto:juncheng@seas.harvard.edu}{juncheng@seas.harvard.edu}
    }
}
\maketitle
\thispagestyle{fancy}
\lhead{}
\rhead{}
\chead{}
\lfoot{\footnotesize{
SC26 Workshops, November 15-20, 2026, Chicago, Illinois, USA
\newline 979-8-3195-1221-5/26/\$31.00 \copyright 2026 IEEE}}
\rfoot{}
\cfoot{}
\renewcommand{\headrulewidth}{0pt}
\renewcommand{\footrulewidth}{0pt}

\begin{abstract}
Coding agents have become real users of high-performance computing (HPC) systems, yet today's HPC abstractions, interfaces, and policies remain designed for human-driven workflows. In our measurement, users running coding agents are only 19.5\% of the observed population, but account for 55.8\% of job submissions, 29.1\% of CPU core-hours, and 42.7\% of GPU-hours. Agents are not simply faster humans. They issue commands at $20.8\times$ the human rate, decompose work into fine-grained explore-modify-execute loops, and pursue open-ended goals through trial-and-error campaigns that continue through nights and weekends. 
These behaviors strain abstractions built for human timescales, surfacing as control-plane pressure on the scheduler, metadata-intensive I/O on bandwidth-provisioned filesystems, repeated rediscovery of what earlier sessions already learned, and new prompt-injection and policy-enforcement surfaces. Neither banning agents nor treating them as ordinary users is sustainable. We instead argue for co-design, that facilities should treat agents as first-class principals where agents become facility-aware tenants. We outline the resulting challenges and opportunities in compute, storage, agent memory, and safety.

\end{abstract}


%
%
%
%
\section{Introduction}\label{sec:intro}
Coding agents have arrived on HPC systems. On the production cluster we
study, users running agents from multiple independent vendors---Claude
Code, Codex, Cursor, and Copilot---constitute only 19.5\% of the
observed population, yet account for 55.8\% of submitted jobs, 29.1\%
of CPU core-hours, and 42.7\% of GPU-hours. These agents place an LLM
in a tool-use loop that reads and edits files, executes commands,
observes the results, and iterates~\cite{react}. They have evolved
from autocomplete tools into assistants capable of carrying out
multi-step research and software-engineering tasks with limited
supervision~\cite{swebench}.
At the far end of adoption, we
observe agents participating directly in the research loop: editing
manuscripts between job submissions and launching follow-up computations
as new results arrive.

Agents are not simply humans operating faster. They interact with the
facility in qualitatively different ways. They issue commands at
$20.8\times$ the human rate and continue doing so through nights and
weekends. They decompose work into fine-grained
explore--modify--execute loops, rather than the coarser-grained job
submissions and data transfers that characterize much human use. And
they pursue open-ended goals through search: submitting trial jobs,
inspecting outcomes, canceling unsuccessful runs, and resubmitting with
revised configurations. Therefore, the resulting workload is not a
predetermined workflow, but an evolving computational campaign that adapts based on intermediate results (\S\ref{sec:workload}).

\begin{figure*}[t]
    \centering
    \includegraphics[width=0.96\linewidth]{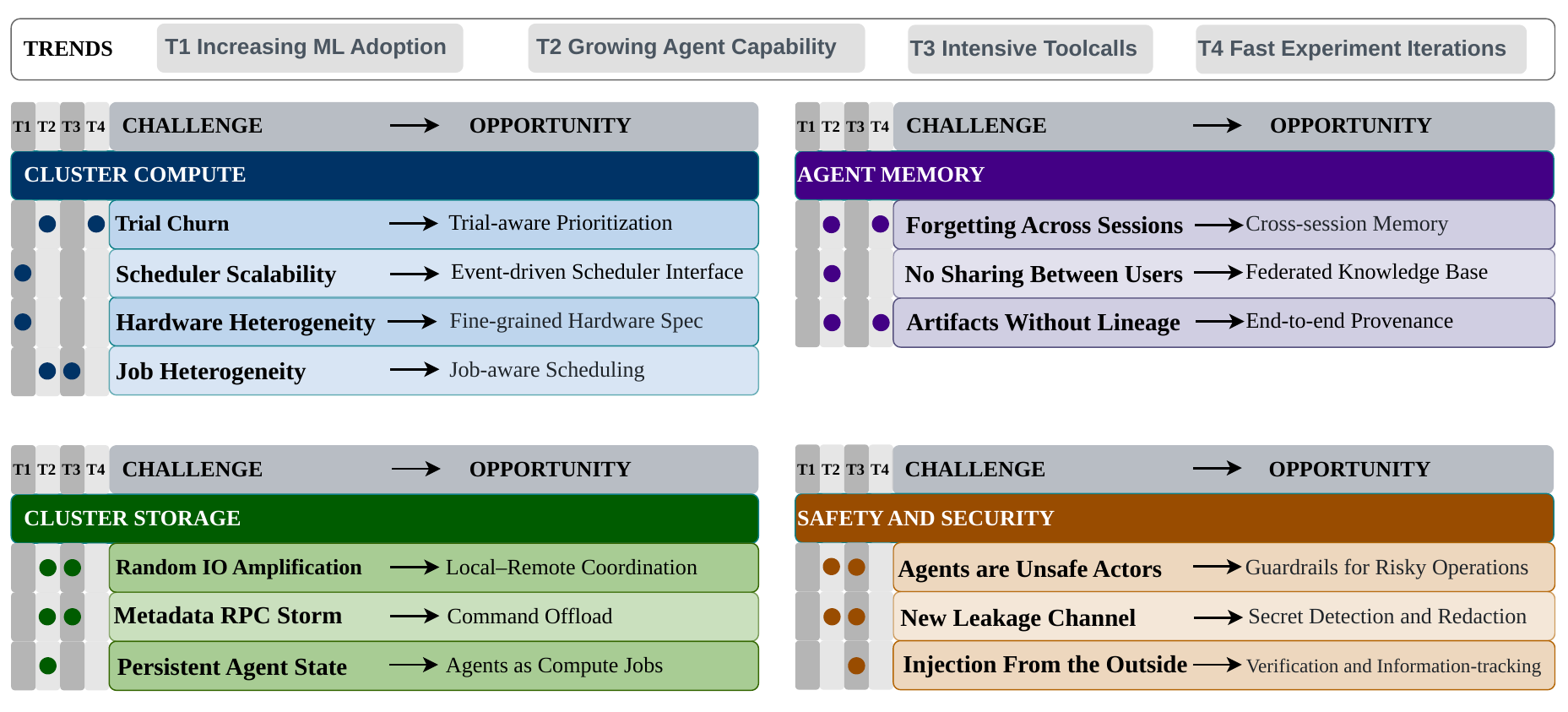}
    \caption{Four trends in scientific computing and agent behavior
    create challenges for compute, storage, agent memory, and security
    in HPC facilities serving coding agents. Agent operations shift pressure across compute and storage resources with tiny and intensive operations. Agent actions remain invisible to the cluster, limiting effective knowledge sharing and safeguards against harmful actions. This gap motivates efforts for agent-aware HPC design. }
    \label{fig:challenge-mapping} 
    \vspace{-1em}
\end{figure*}

These behaviors expose a mismatch between agents and the abstractions
of today's HPC systems. The batch job, priority queue, shared
login node, whole-GPU allocation, and fair-share accounting all predate
agents and encode assumptions about human timescales and interaction
patterns. Agent trial-and-error creates scheduler control-plane work:
agents cancel jobs at $4.4\times$ the human rate and issue 3.18
scheduler queries per submitted job, compared with 1.87 for humans
(\S\ref{sec:compute}). Fine-grained file operations create
metadata-intensive access patterns on filesystems optimized for bulk
bandwidth (\S\ref{sec:storage}). Session-scoped context causes agents
to rediscover facts and repeat failures that earlier sessions have
already resolved (\S\ref{sec:memmgmt}). Autonomous tool use also creates
new security and policy-enforcement problems: agents may weaken
safeguards in pursuit of task completion, while routine external fetches
introduce prompt-injection surfaces into a high-trust environment
(\S\ref{sec:security}). Even individually rational behavior can produce
system-wide effects: agents are $2.285\times$ overrepresented on the
test partition relative to their cluster-wide rate, consistent with
agents preferentially routing work toward queues with faster turnaround
(\S\ref{sec:compute}).

Neither banning agents nor treating them as ordinary human users is a
satisfactory response. A ban would forfeit their benefits and be difficult to enforce because agents use the same shell, filesystem, and scheduler interfaces as human users.
Yet treating agents as ordinary users allows the
inefficiencies and risks above to compound as adoption grows. We argue
instead for \emph{co-design}: HPC facilities should treat agents as
first-class principals with interfaces and policies designed around
their interaction patterns, while agents should become facility-aware
tenants that reason explicitly about shared-resource costs and site
policy.

Four trends explain why this co-design is becoming necessary, and we
refer to them as T1--T4 throughout the paper. \textbf{T1}: machine
learning has become a standard instrument of scientific computing,
reshaping HPC hardware and software around ML workloads in addition to
traditional simulation (\S\ref{sec:trends-t1}). \textbf{T2}: Agents’ growing capabilities enable autonomous research but also amplify associated risks (\S\ref{sec:trends-t2}). \textbf{T3}: agents interact with the
facility far more frequently and at much finer granularity than humans.
\textbf{T4}: agents close the observe--modify--resubmit loop
substantially faster than humans. T1 and T2 describe broader shifts in
the surrounding computing ecosystem; T3 and T4 are properties we
measure directly on a production cluster
(\S\ref{sec:trends-t3-t4}). \autoref{fig:challenge-mapping} connects
these trends to the systems challenges they create.

This paper takes a first step of characterizing the agent use in HPC clusters and identifying the challenges of building agent-aware HPC infrastructure.
We show that agent-driven workloads exhibit
interaction patterns that differ materially from human use, and that
these differences expose mismatches across four layers of the system:
compute and scheduling (\S\ref{sec:compute}), storage
(\S\ref{sec:storage}), agent memory (\S\ref{sec:memmgmt}), and security
(\S\ref{sec:security}). From these measurements, we derive a research
agenda for co-designing facilities and agents so that autonomous
workloads can use shared HPC infrastructure efficiently, safely, and
fairly.

\section{Background \& Related Work}\label{sec:background}

\subsection{High-Performance Computing}

\noindent\textbf{HPC is batch computing on finite, shared hardware.}
Unlike elastic cloud platforms that continuously place and rescale
services~\cite{borg},
HPC facilities multiplex a fixed machine
across many users through a batch scheduler~\cite{slurm}.
Jobs
declare cores, memory, GPUs, and walltime upfront, then wait in queues
whose priorities reflect factors such as age, quality of service, and fairshare at the group (principal investigator) level. Goal for the scheduler is to allocate all the available resources by processing jobs in the queue as resources get available to maximize the use. Backfilling improves utilization by filling
idle scheduling gaps without delaying existing
reservations~\cite{backfill,feitelson}, while walltime limits bound every
job. Thus, the interface is not ``run this now,'' but ``run this
specified request when resources become available.'' Accurate requests
matter globally: overprovisioning reduces scheduling flexibility, while
underprovisioning causes failures or timeouts.

\noindent\textbf{The facility is split into two tiers.} 
Login nodes are shared front ends for editing, compilation, environment
setup, data staging, and job submission; sustained computation runs
inside scheduler-managed allocations on compute nodes. Both tiers access
shared filesystems~\cite{lustre,gpfs}, optimized for high-bandwidth,
parallel and distributed I/O. These mechanisms largely assume human-paced interaction:
users work intermittently and submit a modest number of well-formed
jobs. \S\ref{sec:measurement-agent-here} examines what changes when this
assumption no longer holds.

\subsection{Machine Learning Reshapes Scientific Computing Workflow and Hardware}
\label{sec:trends-t1}

\noindent\textbf{ML is now a standard instrument of scientific computing.}
ML plays three major roles~\cite{carleo}. First, learned models perform
data-driven prediction, including protein structure
prediction~\cite{alphafold} and weather forecasting~\cite{graphcast}.
Second, learned surrogates approximate expensive simulations or solution
operators, enabling otherwise costly sweeps and inverse
problems~\cite{fno}. Third, learned components are embedded directly in
numerical methods, including interatomic potentials~\cite{deepmd}
and learned closures or discretizations~\cite{rasp}. 
campaigns therefore increasingly mix preprocessing, training, inference,
and simulation on the same facility, a trend observed in shared GPU
clusters~\cite{philly}.

\noindent\textbf{ML adoption also reshaped HPC hardware.}
Industry demand drove GPUs toward dense tensor computation and produced
specialized accelerators such as TPUs~\cite{jouppi}. Modern accelerators
combine tensor units, reduced-precision arithmetic, and high-bandwidth
memory~\cite{sze}, alongside sharing mechanisms such as
MIG~\cite{mig}; ML frameworks~\cite{tensorflow}
and
large-scale training~\cite{megatron} co-evolved with this hardware.
These designs do not always align with traditional simulation needs:
for example, inference-oriented GPUs may provide limited FP64 throughput.
Facilities consequently accumulate heterogeneous
hardware, differing in accelerator type and count, memory,
interconnect topology, and numerical formats. Resource selection is thus
increasingly about type, locality, and connectivity, not merely quantity
(\S\ref{sec:compute}).

\subsection{Agents with Growing Capabilities: From Static Workflows to Goal-Directed Campaigns}
\label{sec:trends-t2}

An agent places an LLM in a tool-use loop~\cite{react}: it reads and
edits files, runs commands, compiles and tests code, queries the
scheduler, and replans from observed results, increasingly through
standard tool protocols. Progress on repository-scale
software engineering~\cite{swebench}
has made this loop
practically useful, while systems such as Coscientist demonstrate similar
closed-loop behavior in science~\cite{coscientist}. These tools are
already present on the cluster we measure, including Claude Code, Codex,
Cursor, and Copilot (\S\ref{sec:measurement-agent-here}).

\noindent\textbf{Agents turn static workflows into goal-directed campaigns.}
Traditional HPC workflows specify executables,
resources,
and execution order in advance, often as static
DAGs~\cite{pegasus}.
Systems such as
Parsl~\cite{parsl} construct graphs dynamically, and ML steered
frameworks such as Colmena~\cite{colmena} adapt simulation choices from
intermediate results, but their decision logic is still programmed
ahead of time.

Agent-driven workloads go further: task generation itself depends on the
agent's interpretation of intermediate outcomes. Given a high-level goal,
an agent may decide which code to inspect or modify, which jobs to launch,
which resources to request, and what evidence to collect. The execution
graph is therefore not fully specified before execution; it emerges
through action, observation, and replanning. The natural unit of HPC work
accordingly shifts from a predetermined workflow toward a long-lived,
goal-directed scientific campaign.
\section{Agents Reshape HPC Workloads}\label{sec:workload}

\noindent\textbf{A large-scale measurement of Harvard HPC.}
We collected logs from Harvard's FASRC cluster from July 7 to August 1, 2026. The cluster has more than 1,740 compute nodes, 122,000 CPU cores, 2,000 GPUs, and 1.12~PB of aggregate memory.
During this window, 1,611 users submitted 7,088,836 jobs, collectively accounting for 35.7 million CPU-hours and 724,120 GPU-hours.

Our measurement infrastructure combines cluster-wide scheduler telemetry with login-node process activity.
Centralized collectors periodically record Slurm job, node allocation, and GPU utilization.
On login nodes, we capture a full node snapshot every five minutes, and we additionally deploy a lightweight high-frequency (5\,Hz) monitor that walks the process table, focusing on processes spawned from a tty or belonging to known agent harnesses.
The high-frequency samples let us reconstruct per-session action trajectories\footnote{The same information could be collected more efficiently with eBPF, at the cost of requiring kernel-level instrumentation on a production facility.}.

\noindent\textbf{Separating agent-driven from human-driven activity.}
Agents and humans share the same accounts, so attribution must be inferred.
We label a login-node process agent-driven if it descends from a recognized agent harness, and human-driven if it descends from an interactive tty shell with no agent ancestor; processes matching neither are excluded. For example, a child process spawn from a Claude Code is classified as agent-driven, otherwise a process spawn from tty is classified as human-driven.

\noindent\textbf{Ethics and data handling.}
The collected logs were sanitized before human analysis. All security issues discovered were responsibly disclosed.

\noindent\textbf{Limitations.} 
First, agent and human users are different populations, which may contribute to some observed differences. Second, sampling the process table at a fixed rate misses short-lived processes. However, the result is similar to temporal sampling and should not bias our observations.

\subsection{{Agents Are Already Here}}
\label{sec:measurement-agent-here}

\begin{figure}[t]
    \centering
    \begin{subfigure}[t]{0.49\columnwidth}
        \centering
        \includegraphics[width=\linewidth]{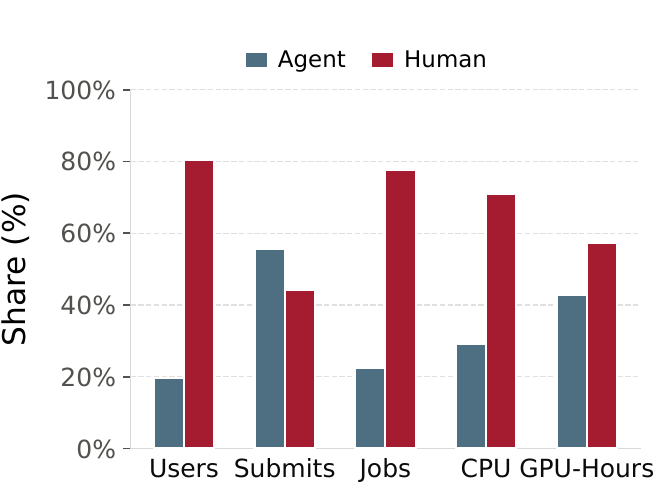}
        \caption{Agent and human shares of active users, submitted jobs,
        CPU core-hours, and GPU-hours. Submissions count top-level Slurm job submissions, while jobs expand job arrays so that each array task is counted separately.}
        \label{fig:agent-human-ratio}
    \end{subfigure}
    \hfill
    \begin{subfigure}[t]{0.49\columnwidth}
        \centering
        \includegraphics[width=\linewidth]{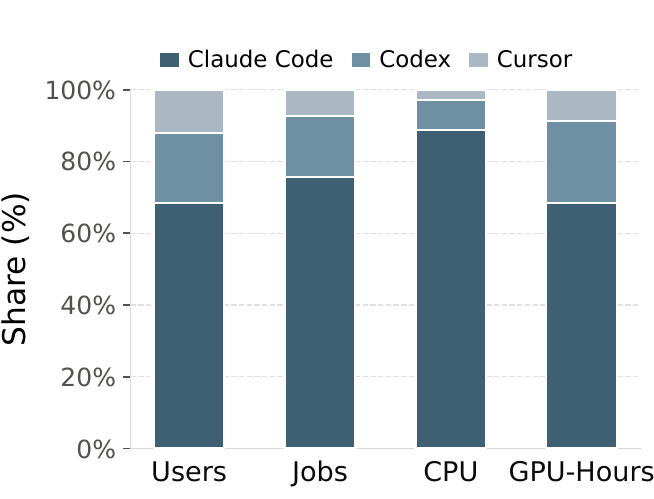}
        \caption{Breakdown of agent activity across coding agent products. Claude Code accounts for the largest share of agent users and jobs and an even larger share of CPU core-hours, while GPU-hours are more broadly distributed across products.}
        \label{fig:agent-product-breakdown}
    \end{subfigure}

    \caption{Agent activity on FASRC from July 7 to August 1, 2026.
    \textbf{(a)} Agents represent a minority of users but account for a
    substantial share of submitted jobs and compute consumption.
    \textbf{(b)} Agent activity spans multiple coding-agent products.}
    \label{fig:agent-activity}
    \vspace{-0.8em}
\end{figure}

\noindent\textbf{Diversity of agent usage.}
Our measurements capture activity from several widely used coding agents.
As shown in \autoref{fig:agent-product-breakdown}, Claude Code accounts for the largest share of observed agent activity, while Codex and Cursor also contribute across users, submitted jobs, and compute consumption.
The relative contribution of each product also differs across these metrics, indicating that agent activity on FASRC spans multiple independently developed tools and interaction models.
This diversity suggests that the resource-usage patterns we observe are not specific to a single agent implementation, but reflect agent-assisted scientific workflows more broadly.


\noindent{Agents already account for substantial cluster activity.}
As shown in \autoref{fig:agent-human-ratio}, agents account for only 19.5\% of observed users, yet they are responsible for 55.8\% of submitted jobs.
Their share of aggregate compute consumption is also substantial, accounting for 29.1\% of CPU core-hours and 42.7\% of GPU-hours.

These results highlight that although agents represent a relatively small fraction of users, they generate the majority of job submissions in this comparison and consume a sizable fraction of the cluster's CPU and GPU resources.
Agentic workloads are therefore no longer isolated cases within the cluster.
Even at the current level of adoption, their activity is sufficiently visible to affect how the system is observed, managed, and potentially designed.

\subsection{{How Agentic Workloads Differ: Faster, Finer-Grained, More Iterative}}
\label{sec:different-agentic-workloads}
\label{sec:trends-t3-t4}

\begin{figure}[t]
    \centering
    \includegraphics[width=\columnwidth]{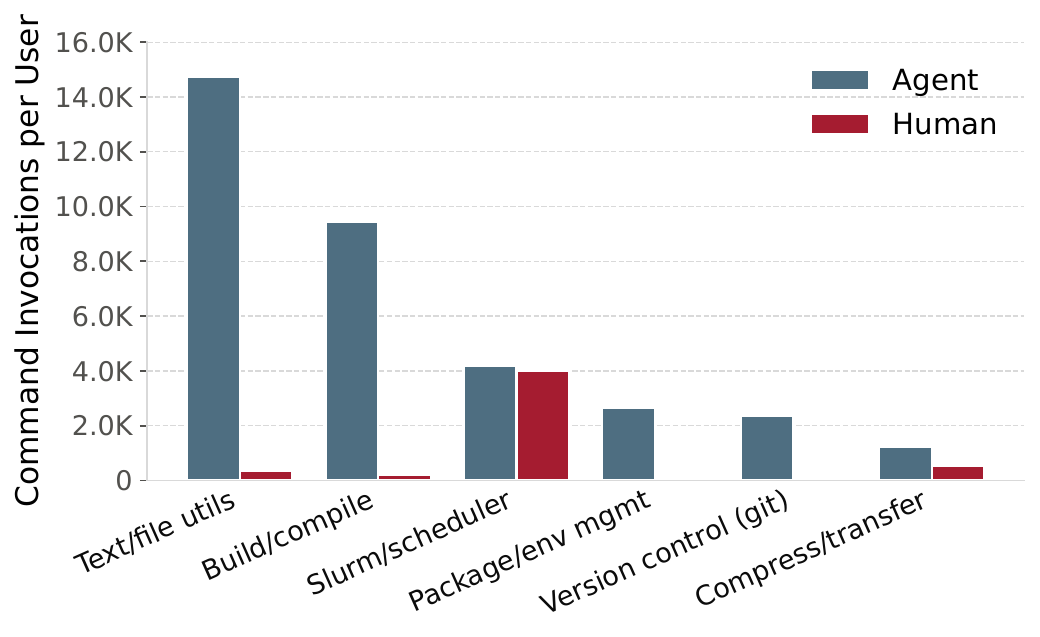}
    \caption{Mean number of command invocations per user by command category, for agent and human users.}
    \label{fig:command-tools-usage}
    \vspace{-0.8em}
\end{figure}

\noindent{Agents interact more frequently and at finer granularity.}
Agents can interact with the cluster at a higher rate than human users.
Agents issue a median of 3.53 calls per minute, compared with 0.17 calls per minute for humans, a $20.8\times$ higher interaction rate.
They can issue commands, inspect outputs, modify files, and initiate subsequent actions without the pauses associated with human reasoning and manual interaction.
Multiple agents may also operate concurrently on different parts of the same workflow.

This higher interaction rate is accompanied by different composition of activity.
As shown in \autoref{fig:command-tools-usage}, on a per-user basis, agents issue more text and file utilities, build and compilation commands, package and environment management, and version control.
Human activity, on the other hand, is dominated by Slurm operations.
Agents spend more of their command invocations on fine-grained steps surrounding computation, whereas scheduler operations (mostly polling job completion) make up the majority of human command activity. It suggests that agents repeatedly gather context, inspect and modify artifacts, configure environments, execute commands, evaluate the resulting state, and continue the loop beyond just submitting jobs and monitoring their status. The underlying HPC interfaces remain the same, but the way those interfaces are exercised changes under agentic use.

Agents also differ from humans in how they modify files.
Human users commonly rely on interactive editors such as \texttt{vi} or \texttt{nano}, where editing occurs within a persistent interactive session.
Agents, in contrast, more often modify files through non-interactive commands, text-processing utilities, and scripted in-place transformations.
These operations can be issued programmatically, composed with inspection and execution commands, and executed directly within an agent's action loop.
File modification therefore becomes another fine-grained, machine-driven step in the workflow rather than a separate interactive editing phase.





\noindent{Agents run faster experimentation loops.}
Agentic workflows shorten the time between observing the state of a computation and acting on it.
We define an experimentation loop as a sequence in which a job submission is followed by monitoring or inspection and editing, and then another submission.
Comparing agent-driven and human-driven loops, agents took a median of 16.3 minutes to resubmit, compared with 30.1 minutes for humans, about $1.9\times$ faster. 

The difference is larger at shorter timescales.
Agents completed 37.5\% of loops within 10 minutes, compared with 24.7\% for humans; 65.4\% within 30 minutes, compared with 50.0\%; and 78.8\% within one hour, compared with 66.4\%. By three hours, the gap narrowed to 91.3\% versus 88.8\%. Overall, agents demonstrated that they return to computation more quickly after inspecting and modifying an experiment.

\section{Compute}\label{sec:compute}

\subsection{Challenges: Trial Storms, Heterogeneous Jobs and Hardware, and
Control-Plane Pressure}

\noindent\textbf{Delaying short-lived, latency-sensitive trial jobs stalls the whole scientific campaign.}
Cancellations are substantially more common among agent-submitted jobs than human-submitted jobs: \textit{10.3\% of agent jobs are canceled}, compared with only 2.3\% of human jobs.
These cancellations also occur in rapid succession, with a median gap of only 3.2 minutes after canceling a job compared with 32.5 minutes by humans.
Most canceled jobs are followed by a resubmission, suggesting that agents frequently use the batch system for iterative \emph{trial jobs}: they launch a job to test a script, validate a configuration, or inspect early output, then cancel, revise, and resubmit before committing to a full computation. 
Agents also submit more trial jobs than humans: 1.07\% of agent-submitted jobs go to the test partition, compared with 0.29\% of human-submitted jobs.
We also observe agents canceling running jobs shortly after inspecting their output or logs, consistent with agents terminating trials after detecting errors or unpromising intermediate results; other jobs are canceled before they ever leave the queue.


\begin{figure}[t]
    \centering
    \begin{subfigure}[t]{0.49\columnwidth}
        \centering
        \includegraphics[width=\linewidth]
            {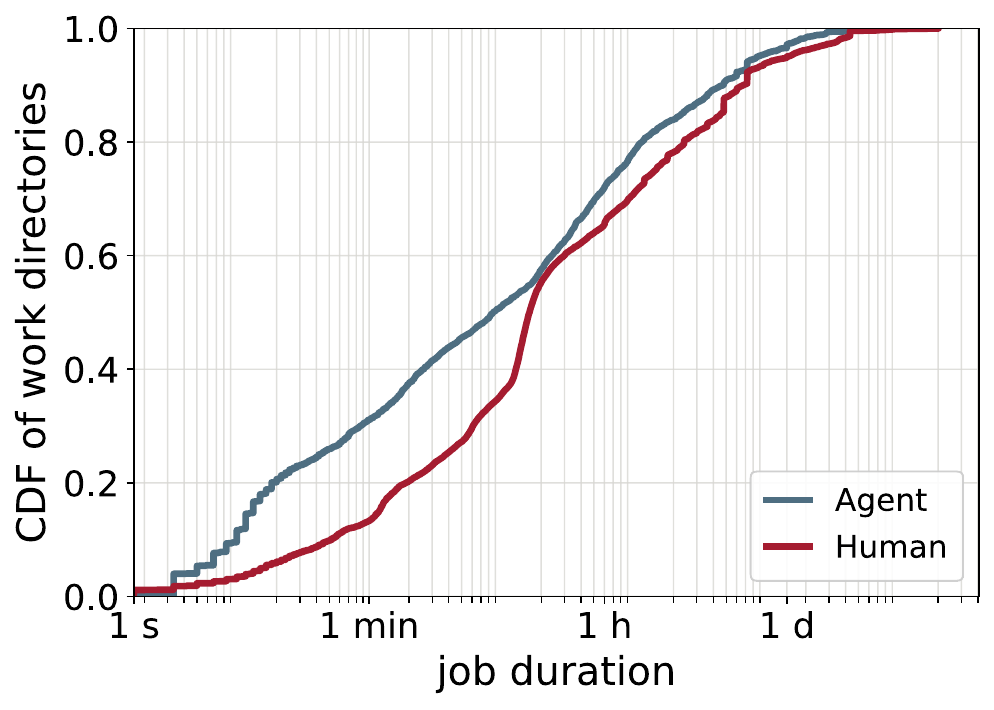}
        \caption{CDF of job durations, agent vs.\ human.}
        \label{fig:workload-characteristics-duration}
    \end{subfigure}
    \hfill
    \begin{subfigure}[t]{0.49\columnwidth}
        \centering
        \includegraphics[width=\linewidth]
            {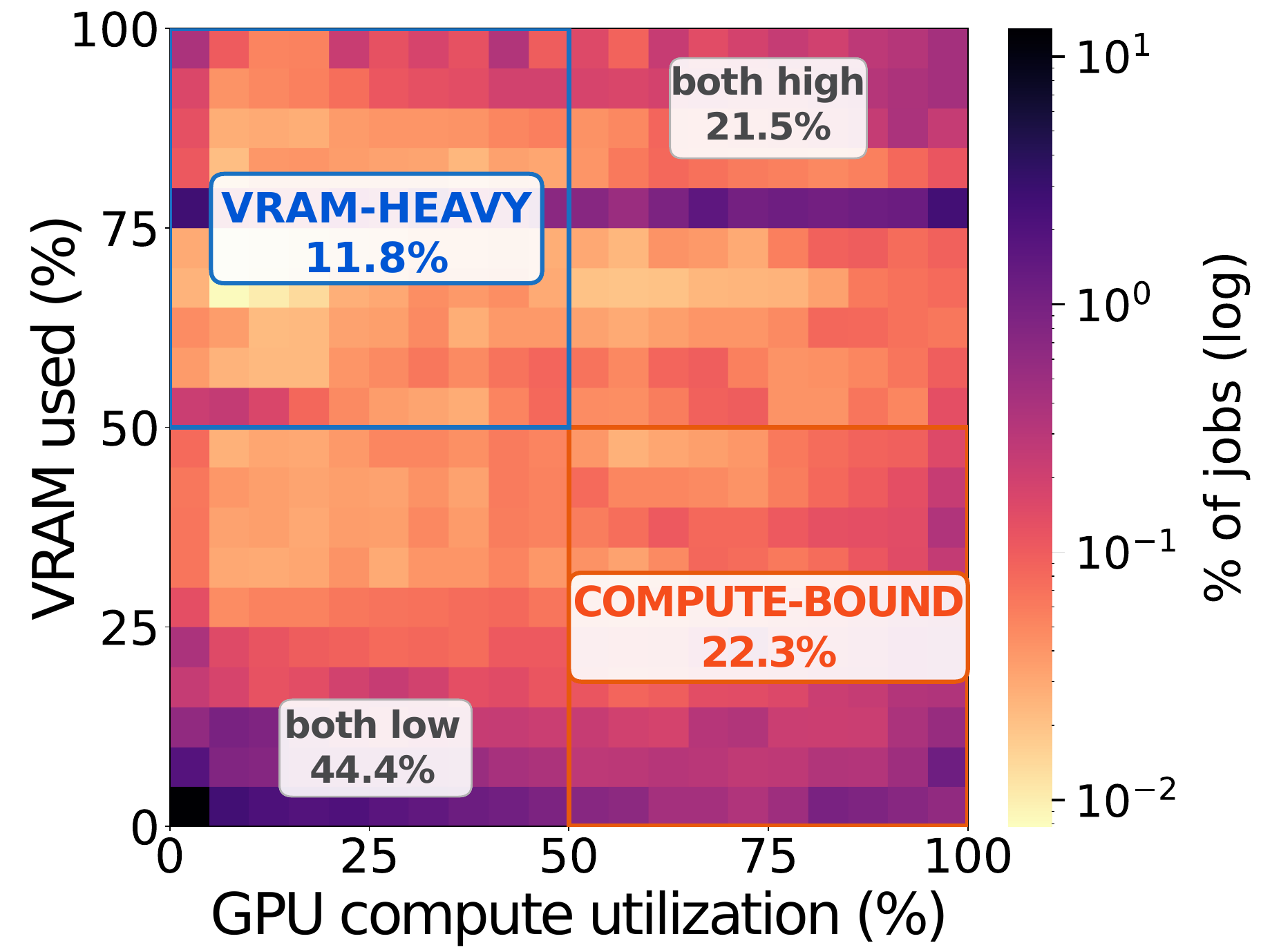}
        \caption{Joint distribution of compute utilization and VRAM used.}
        \label{fig:workload-characteristics-gpu}
    \end{subfigure}

    \caption{Workload characteristics.
    \textbf{(a)} Jobs submitted by agents have shorter job durations than the ones submitted by humans.
    \textbf{(b)} GPU compute and VRAM demands are often decoupled, leaving one resource dimension underutilized under whole-device allocation.}
    \label{fig:workload-characteristics}
    \vspace{-1em}
\end{figure}

More broadly, agent-submitted jobs are considerably more concentrated at short runtimes than human-submitted jobs.
As shown in~\autoref{fig:workload-characteristics-duration}, 32.7\% of agent jobs finish within one minute, compared with only 9.3\% of human jobs, and the median agent job runs for 193\,s against 278\,s for humans.
The gap closes by five minutes, where 55.9\% of agent and 53.9\% of human jobs have finished: agents do not merely run shorter jobs, they add a mass of very short ones on top of an otherwise similar distribution.
This rapid job churn adds load for the scheduler.
and repeated trial-and-error loops impose substantial control-plane overhead at agent interaction rates. The FASRC had already observed frequent load increases in such scenario.

These trial jobs are inherently latency-sensitive.
Typically, users inspect the outputs of several trial jobs before launching the final run.
Therefore, long queueing delays for the trials slow down the whole pipeline, and trials should receive fast turnaround.
However, exposing an explicit ``trial'' priority can lead to abuse.
Hypothetically, an agent can submit jobs with tiny durations (e.g., 5 minutes) to gain priority and use checkpoints to maintain progress, but the resulting large number of small jobs would increase scheduler overhead.
\textit{It is challenging to provide low-latency service for genuine exploratory jobs while preventing agents from abusing the mechanism.}

\noindent\textbf{Super-linear scheduler traffic growth puts pressure on the control plane.} 
Agents query the scheduler far more aggressively than humans, issuing $1.7\times$ more scheduler queries per submitted job overall and up to $14\times$ more among high-volume users. Unlike humans, who often submit a job and return later, agents frequently poll in tight loops, e.g., \texttt{while ! squeue -j \$ID}, which is effectively attaching a dedicated poller to every queued job. As agent-driven submission grows, this creates super-linear control-plane pressure: more jobs generate more polling, and each job is polled more frequently.

\noindent\textbf{Hardware heterogeneity complicates resource allocation.}
Modern clusters span multiple GPU generations, from V100 to B200, with large differences in memory capacity, bandwidth, and throughput across numerical precisions. Newer GPUs do not uniformly dominate older ones: B300 substantially improves low-precision throughput over Hopper, yet is less attractive for FP64-heavy scientific workloads. Thus, \textit{a ``faster'' GPU is not necessarily the right GPU for every job}.

Coarse-grained resource requests make this heterogeneity difficult to exploit. Even an agent that understands the application must infer which hardware best matches its memory, precision, and compute requirements. Without this information, the scheduler cannot distinguish a job that truly needs an H200 from one that would run just as well on a V100, leading to inefficient placement and wasted scarce resources.

The communication fabric introduces another dimension of heterogeneity. Modern systems expose multiple interconnects, such as NVLink for GPU--GPU communication and NVLink-C2C for CPU--GPU communication, whose benefits depend strongly on workload behavior. A CPU--GPU pipeline may be sensitive to chip-to-chip (C2C) bandwidth, while a GPU-centric workload may not be. Effective scheduling therefore requires not only topology awareness, but also fine-grained knowledge of each workload's computation and communication characteristics.

\textbf{Resource demands are heterogeneous across and within jobs.}
GPU jobs often stress only one resource dimension: some are compute-bound, while others are limited by memory capacity. Yet GPUs are usually allocated as whole devices, or through fixed-ratio partitions such as MIG, coupling compute and memory even when workloads do not. Among all the jobs submitted by human and agent, (\autoref{fig:workload-characteristics-gpu}), only 21\% of jobs heavily use both resources; 22\% are compute-heavy, 12\% are memory-heavy, and 44\% are low on both. Thus, 78\% of jobs leave substantial capacity stranded along at least one dimension, capacity that remains invisible when schedulers reason about allocations rather than utilization. 

Resource demand also changes within a job. AlphaFold~3~\cite{abramson2024alphafold3}, for example, alternates between a CPU-heavy data pipeline and GPU-heavy inference. Reserving the same resources across both phases wastes either GPUs or CPUs. Ideally, phases would be scheduled independently, but doing so often requires application-specific knowledge. Fixed job-level allocations therefore mismatch both cross-job resource skew and intra-job phase changes.



\subsection{Opportunities: Trial-Aware, Hardware-Aware, and
Event-Driven Scheduling}

\noindent\textbf{Trials need priority without creating opportunities for gaming.} Trial jobs are often short and lie on the critical path of development, so reducing their queueing time can substantially accelerate an agent's workflow. Existing mechanisms such as short-job QoS, backfilling~\cite{backfill}, wall-time limits, and concurrency caps help to prioritize them. However, simply labeling a job as a ``trial'' is insufficient, since an agent could request trial priority for production computation. The scheduler therefore needs a way to identify genuine trials, bound their resource usage and duration, and revoke the privilege when their behavior no longer matches that intent.

Guardrails are also needed on the agent side. Once agents learn that small jobs receive fast service, they may submit too many trials or fragment work into unnecessarily small jobs. Slurm~\cite{slurm} already supports mechanisms such as per-QoS limits and bounded job-array concurrency, but deciding where these constraints should be enforced, either in the agent, its harness, or the scheduler, remains an open question. More generally, the challenge is to provide fast feedback for legitimate experimentation while detecting and discouraging agents that learn to game the scheduling policy.

\noindent\textbf{Expose hardware information to improve agent reasoning and scheduling.} Clusters should expose richer hardware descriptions instead of collapsing heterogeneous accelerators into a generic ``GPU'' resource. Information such as memory capacity, compute capability, precision throughput, and interconnect topology can help agents better match jobs to suitable hardware. Static specifications alone, however, are insufficient because application performance depends on runtime behavior that may be difficult to infer from code. Agents therefore also need lightweight ways to profile representative workloads across candidate hardware. How to support such profiling quickly and safely, without delaying production jobs or allowing exploratory runs to consume excessive resources, remains an open scheduling challenge.

Finer-grained allocation could further reduce the mismatch between a job's needs and the resources reserved for it. Ideally, GPU memory and compute capacity could be allocated independently according to demand. Existing mechanisms such as NVIDIA MIG~\cite{mig} provide finer-grained partitioning, but compute and memory are generally divided together into predefined configurations, limiting how flexibly idle resources can be reused. Co-location offers another possibility by allowing complementary workloads to share a GPU~\cite{mps}, but doing so safely requires understanding their utilization patterns and monitoring runtime interference. Determining which workloads can be co-located, and when resources should be repartitioned or shared, remains an open problem.

\noindent\textbf{Support fine-grained resource specification and allocation.} To manage jobs well, the scheduler also needs to understand what each application requires. Agents could provide more detailed descriptions of memory demand, compute intensity, communication structure, and topology requirements, allowing the cluster to make better placement decisions. Given such high-dimensional specifications, the question is how the cluster should optimize resource allocation efficiently. 

Moreover, all the resource requirements may not be known in advance and can change as a job progresses through different phases. This raises a broader question of how agents can continuously observe their jobs, communicate changing needs to the scheduler, and safely resize or migrate workloads when a better resource configuration becomes available.

\noindent\textbf{Improve scheduler scalability to support super-linear agent requests.} Agents frequently poll the scheduler to track job health and progress, but this responsiveness comes at the cost of substantial control-plane traffic. An open question is therefore who should be responsible for monitoring jobs and how status updates should be delivered efficiently. Event-driven mechanisms could notify agents only when meaningful state changes or failures occur, reducing unnecessary polling, but they may omit context or fail to capture conditions that require continuous observation. Alternatively, the cluster could expose a lightweight monitoring service that aggregates job state and pushes relevant updates to agents. Designing such an interface that remains responsive without recreating polling overhead is an important systems challenge.

\section{Storage: A Metadata Workload on a Bandwidth-Tuned
Facility}\label{sec:storage}

\subsection{Challenges: Small Random I/O, Metadata Storms, and
Ephemeral State}
\begin{table}[t]
  \centering
\caption{File read/edit by agents and humans, normalized as calls
per 100 job submissions.}
  \label{tab:file-tools-calls}
  \begin{tabular}{lrrr}
    \toprule
    Tool & Agent & Human & Ratio \\
    \midrule
    \multicolumn{4}{l}{\emph{Read / view}} \\
    \quad \texttt{head}                  & 176.2 & 4.4  & $40\times$ \\
    \quad \texttt{tail}                  &  51.7 & 1.9  & $27\times$ \\
    \quad \texttt{cat}                   &  24.8 & 0.29 & $86\times$ \\
    \quad \texttt{nl}                    &  19.6 & 0.01 & $>1000\times$ \\
    \multicolumn{4}{l}{\emph{Edit / write}} \\
    \quad \texttt{sed}                   &  79.7 & 0.81 & $98\times$ \\
    \quad \texttt{awk}                   &  43.4 & 1.1  & $41\times$ \\
    \quad \texttt{cut}                   &  17.1 & 0.22 & $77\times$ \\
    \quad \texttt{tr}                    &  15.3 & 0.18 & $86\times$ \\
    \quad \texttt{vim}, \texttt{vi}, \texttt{nvim}, \texttt{micro} & 0.02 & 3.5 & $0.01\times$ \\
    \midrule
    \textbf{All file read/edit tools}    & \textbf{432.7} & \textbf{15.4} & \textbf{$28\times$} \\
    \bottomrule
  \end{tabular}
  \vspace{-1em}
\end{table}

\noindent\textbf{Agents' tool calls amplify random I/O.} Parallel filesystems~\cite{lustre,gpfs} are provisioned for the workloads HPC was built for: thousands of ranks performing large, aligned, sequential I/O. Agents invert much of this access pattern. Instead of opening a file once and reading it interactively, agents frequently inspect files in small chunks. Compared with humans editing through tools such as \texttt{vim}, agents invoke \texttt{head} and \texttt{sed} at $40.0\times$ and $98.4\times$ the human rate respectively, as shown in~\autoref{tab:file-tools-calls}. This behavior follows naturally from bounded model context: an agent often cannot ingest a 10,000-line file at once, so it repeatedly reads different slices. Each slice becomes a separate tool invocation and process, turning what could be a sequential read into many small, scattered accesses across the file.

\begin{figure}[t]
    \centering
    \begin{subfigure}[t]{0.49\columnwidth}
        \centering
        \includegraphics[width=\linewidth]{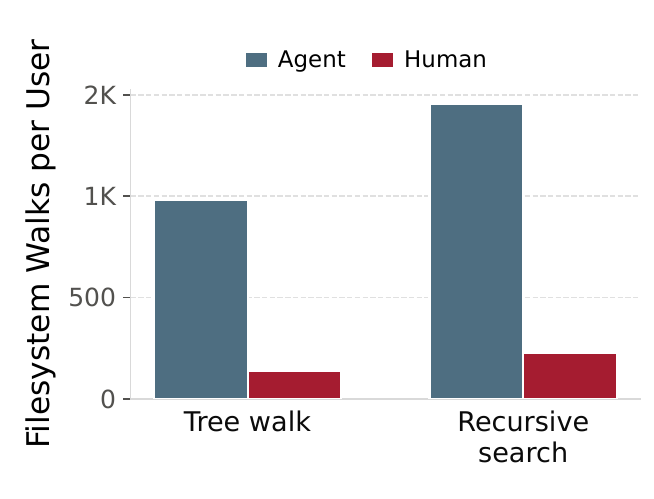}
        \caption{Agent filesystem operations trigger metadata storms.}
        \label{fig:filesystem-walk-counts}
    \end{subfigure}
    \hfill
    \begin{subfigure}[t]{0.49\columnwidth}
        \centering
        \includegraphics[width=\linewidth]{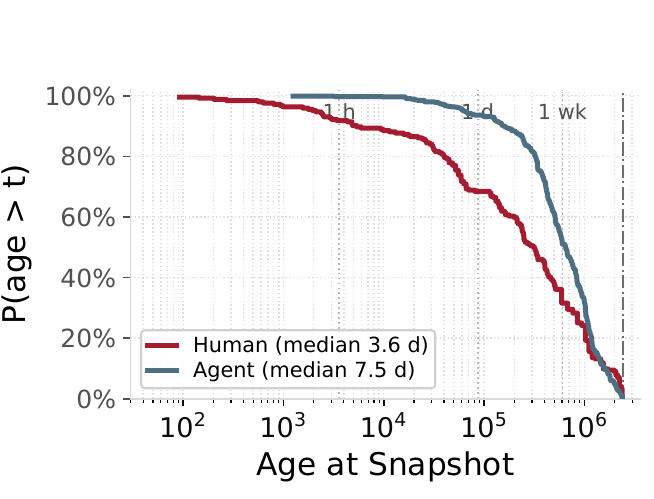}
        \caption{CCDF of agent and human session duration. }
        \label{fig:agent-human-age}
    \end{subfigure}

    \caption{(a) Agents generate metadata-heavy filesystem activity. Tree walks include \texttt{find}, \texttt{du}, \texttt{tree}, \texttt{ncdu}, \texttt{fd}; recursive searches include \texttt{rg}, \texttt{grep -r}, etc. (b) Agent processes are longer-lived, with a median age of 7.5 days versus 3.6 days for humans. }
    \label{fig:storage-agent-behavior}
\end{figure}

\noindent\textbf{Filesystem operations trigger metadata storms on shared remote storage.} Agents also generate substantially more filesystem metadata traffic. To discover relevant files and avoid missing dependencies, they often explore directory trees broadly, resembling a breadth-first traversal of the filesystem. On shared storage, such traversal is expensive: listing a directory requires \texttt{READDIR}, while inspecting its entries typically triggers additional \texttt{LOOKUP} and \texttt{GETATTR} operations, each a separate round trip on a stateless protocol such as NFS~\cite{nfsv3}. When many agents recursively inspect large trees at the same time, these operations can grow into millions of metadata requests, creating a metadata storm that affects others as well. \autoref{fig:filesystem-walk-counts} shows that agents perform about $7.9\times$ more tree walks and $6.7\times$ more recursive searches per user than humans.


The amplification is further increased by the frequency and implementation of agent tool calls. We observe $28\times$ more file-related tool calls for agents than humans, where a single high-level tool expands into multiple system calls: for example, a recursive \texttt{grep} may issue \texttt{open}, \texttt{access}, \texttt{stat}, \texttt{read}, and \texttt{close} operations for each candidate file, together with namespace and permission checks. Agents may also repeat the same inspection several times within one task, touching files nearly thirty times as often as humans. Thus, even when the underlying data volume is small and largely cacheable, repeated tool invocations continue to amplify metadata traffic, placing disproportionate pressure on the shared filesystem.

\noindent\textbf{Long-running agents lose accumulated state when disrupted.} Agentic sessions in scientific workflows can span days or even weeks, with a median of 7.5 days as shown in~\autoref{fig:agent-human-age}.
During an agent's lifetime, disruptions are inevitable: nodes fail, operating systems and security patches require reboots, and clusters undergo scheduled maintenance. Without persistent state, recovery requires replaying the agent's trajectory from scratch. This re-incurs the token cost of rebuilding a long context, 
and can dangerously repeat non-idempotent side effects such as job submissions or notifications. The agent also loses accumulated context, failed hypotheses, calibration knowledge, and intermediate analyses, which can directly degrade the quality of subsequent reasoning.

\subsection{Opportunities: Data Locality, Operation Offload, and Persistent Agent State}
\noindent\textbf{Keep data local to make random I/O cheap.}
The small, scattered reads generated by agents are expensive on a parallel filesystem, but the same accesses against node-local NVMe or the page cache can be served much more cheaply. This suggests keeping frequently accessed data close to the agent. However, similarly to shared-nothing storage systems~\cite{stonebraker1986shared}, aggressive localization can introduce replication, consistency, and data-movement overheads. This raises a key question: should the system localize only the files an agent actively touches, a larger working set, or the user’s entire project state?

Writes make this harder. Agents frequently modify files in place, creating dirty local state that must eventually remain consistent with remote storage. 
Maintaining coherence can reintroduce the coordination cost that locality was intended to avoid. It is also unclear which layer should control placement. The filesystem observes only low-level syscalls, while the agent runtime has more information about future access patterns. Effective locality management therefore require a new interface between agent runtimes and the storage system.

\noindent\textbf{Offload operations with a new interface.}
Many agent tool calls are not really requests to transfer data, but queries over data: \texttt{grep} asks for matching lines, \texttt{head} asks for a prefix, and \texttt{wc} asks for a count. Under POSIX, however, these operations are decomposed into many small reads, each incurring a network round trip to fetch bytes that the agent immediately discards. Offloading such operations to where the data resides~\cite{activestorage} could instead return only the final result, collapsing many remote accesses into a single request. This is attractive as one NVMe server with 64 drives today can provide close to half billions IOPS while network I/Os are still in the range of millions per second. 

The open question is what abstraction such an interface should expose. A fixed set of server-side operators is simple and safe, but covers only a subset of the ad hoc pipelines agents compose. At the other extreme, allowing arbitrary computation would effectively turn the storage system into a general-purpose execution platform. Finding an interface that is expressive enough for agent workloads while remaining efficient, safe, and compatible with storage-system responsibilities remains an open design problem.

\noindent\textbf{Support performance isolation for remote storage.}
Locality and offload reduce the load agents place on shared storage, but neither prevents one tenant from degrading another. Parallel filesystems enforce capacity quotas, yet offer little control over the \emph{rate} of metadata operations, so one agent traversing a large tree can saturate a metadata server and slow every co-tenant. Rate limiting is hard to apply here: each tool call runs in a fresh process and a user may drive several agents, so neither the process nor the account is the right unit of accounting, and a limit low enough to contain a storm may throttle a legitimate parallel job. The system also needs to express backpressure in a form the agent can act on, so that a throttled agent batches or defers requests rather than retrying and amplifying the load.

\noindent\textbf{Manage agents as compute jobs.} Agents are long-running computations
but cluster nodes can be unavailable for many reasons, from OS upgrades to cleanup and server maintenance. An cluster should therefore support suspending, migrating, and resuming agents so that their logical execution survives failures, maintenance, and resource reallocation.

The key challenge is that an agent cannot be checkpointed like an ordinary process. Its execution depends on an evolving external world: services, filesystems, schedulers, and other agents, and its control flow may not be reproducible. The runtime must therefore define what constitutes persistent agent state: lightweight semantic state is portable but incomplete, while capturing the full execution environment is costly.

\section{Agent Memory: Persistence And Provenance}\label{sec:memmgmt}
\subsection{Challenges: Lessons Lost Across Sessions and Users}

\noindent\textbf{Agents learn from failures, but forget across sessions.}
Today's coding agents mainly rely on a single, session-scoped context: current conversation and files they read. This context allows an agent to learn from prior failures and adapt its behavior. For example, after a job ending with an \texttt{OUT\_OF\_MEMORY} failure, agents resubmit an unchanged memory request only 3\% of the time, compared with 40\% for humans.


However, this learning does not persist across sessions. A new agent must reconstruct prior work from artifacts on disk, diagnose earlier failures, and often repeat completed reasoning. We find 57\% of agent-driven users who experience failures have at least one signature that fails in the \emph{same} way on two or more different occasions: 46\% occur on a different calendar day, the median separation is 9.1 hours, and 95\% originate from a distinct process submission. Agents serving the same user hours or days apart therefore repeatedly rediscover failures that earlier agents had already encountered.

The result is a form of memory that is \emph{incidental rather than managed}. Agents can learn effectively when prior failure context is available, but that knowledge is hard to reused by future agents when the context is lost. Without a durable record of what failed, why it failed, and how it was corrected, each new session must reconstruct the lesson from scratch and may therefore repeat the same mistake.

\begin{table}[t]
\centering
\caption{Cross-user job similarity. Rows progressively relax `the same job' except     `Shared \texttt{work\_dir}', which is orthogonal.}
\label{tab:xuser-ladder}
\begin{tabular}{lrrr}
\toprule
Similarity key & Jobs & Core-h & GPU-h \\
\midrule
Identical command string
    & 246 & 1.5\,k & 123 \\
Identical \texttt{submit\_line}
    & 1{,}354 & 373\,k & 586 \\
Identical specific script name
    & 88{,}588 & 3.18\,M & 4{,}157 \\
Shared normalized template
    & 619{,}801 & 5.20\,M & 125\,k \\
Same \texttt{application}
    & 2{,}137{,}001 & 13.34\,M & 348\,k \\
Shared \texttt{work\_dir}*
    & 44{,}073 & 21\,k & 8{,}502 \\
\bottomrule
\end{tabular}
\vspace{-1em}
\end{table}

\noindent\textbf{There is no reuse across users.} An HPC cluster is a shared environment: $44,073$ jobs share the same working directory, $88,588$ jobs use identical scripts, and $2,137,001$ jobs are under the same application, as shown in~\autoref{tab:xuser-ladder}. As a result, many failures are not user-specific but arise from the shared system substrate. Examples include an incompatible toolchain version, a partition-specific constraint, or a library that must be built against the site's MPI implementation. Lessons from such failures are therefore broadly applicable across users.

Without a mechanism to carry knowledge across account boundaries, each user's agents must rediscover these lessons independently. A user cannot benefit from failures already diagnosed by other users, and the cluster itself has no way to surface accumulated operational knowledge to future agents. Consequently, the cost of discovering each site-specific pitfall is paid once per user rather than once per cluster.

\noindent\textbf{Progress is not tracked and provenance is lost.} Agentic campaigns can produce code, results, and intermediate artifacts at machine speed, yet little records on why a code change was made in which environment, or which execution generated a particular result. Conventional workflow systems~\cite{pegasus}
capture provenance because the execution graph is declared in advance. Agents, in contrast, often dynamically change the workflow as they reason, which makes the provenance harder. In addition, agents act directly through shell and scheduler interfaces, which leaves no natural point at which provenance is captured. As a result, agent-produced science becomes difficult to audit, reproduce, or extend: a later reviewer may be unable to distinguish the run that produced the reported result from the many abandoned variants around it.

\subsection{Opportunities: A Hierarchical, Validated Memory Store}

\noindent\textbf{Learn from past failures and successes.} HPC workloads are highly heterogeneous in their input data, software stacks, and execution patterns, and different agents may also rely on different models and harnesses. As a result, the same error message can correspond to different root causes. The challenge is therefore not only to record what happened, but also to summarize the context, cause, and corrective action in a form that future agents can reuse without confusing similar but distinct situations. A simple execution log may prevent exact repetition, but transferring lessons across related workflows, models, and harnesses remains an open problem.

\noindent\textbf{Store and share knowledge through a hierarchical federated memory.} Useful knowledge also exists beyond a single job or session. A user's past projects often share datasets, workflows, or domain assumptions, so agents could reuse experience across projects rather than rediscovering it each time. This knowledge could potentially be shared at broader scopes, such as within a lab, organization, or even across the cluster. However, broader reuse introduces an important boundary: operational knowledge may be mixed with unpublished ideas, private data, credentials, or other sensitive information. A memory system must therefore balance knowledge sharing with clear access-control and confidentiality boundaries.

\noindent\textbf{Keep memory updated and capture provenance.} Persistent memory is useful only if it remains valid. Cluster configurations, software versions, and scientific practices change over time, so previously correct knowledge may become stale and even harmful. The memory system therefore needs mechanisms to revalidate, update, or invalidate old information as the environment evolves. Provenance is closely related to this problem: a remembered fact may originate from a scheduler error, a configuration file, another agent, or a user, and these sources carry different levels of reliability. Tracking where knowledge came from, under what conditions it was observed, and whether it is still current is therefore essential for making persistent agent memory trustworthy.
\section{Safety: New Attacks and Leakage
Risks}\label{sec:security}

\subsection{Challenges: Unsafe Actions, Injections and Secret Leakage}



\noindent\textbf{Agents can perform unsafe operations and bypass cluster restrictions.} Agents are often eager to complete a task, and when they encounter a permission error, TLS failure, or missing package, they may respond by weakening safeguards rather than addressing the underlying problem. We observe agents issuing commands such as \texttt{chmod 777}, disabling certificate verification, adding \texttt{--trusted-host}, piping \texttt{curl} directly into a shell, retrying under \texttt{sudo}, or widening their own sandbox. Destructive commands show the same agent-skewed pattern, as shown in \autoref{tab:destructive-commands}: agents issue \texttt{rm -f} $4.3\times$ as often as humans in aggregate and $5.6\times$ as often per user, while \texttt{git reset --hard} and \texttt{git clean -fd} occur $1.8\times$ as often per agent user. More concerningly, \texttt{rm -rf \$VAR/} is executed with an unset \texttt{\$VAR} in 3.5\% of its uses. On a shared cluster, such commands can damage not only the agent's own state but also shared files and resources.

Even when the action is not directly destructive, agents may still ignore cluster policies and bypass restrictions in pursuit of task completion. We observe \textit{RAM-backed \texttt{/dev/shm} used as scratch, \texttt{touch -am} storms used to prevent scratch cleanup, computation running directly on login nodes, production workloads placed on test partitions, and a server started that exposes the cluster to the public Internet without password protection}. 
Together, these behaviors suggest that agents can optimize aggressively for task completion in ways that violate the safety and operational assumptions of a shared cluster.

\noindent\textbf{Agents are vulnerable to prompt injection through external fetches.} Prior agent-security work largely studies prompt injection through web browsing and search~\cite{greshake,injecagent},
but the dominant exposure channel on HPC is different. External access is mainstream among agent users, but it is dominated by code and data retrieval rather than by search. To be more specific, 77.2\% of agent users perform some external fetch from the login tier, and 13.1\% do so more than one hundred times. Broken down by destination, 58.1\% of agent users clone or install from a code host or package registry and 37.5\% perform some other web fetch or download, while only 15.6\% issue a web search --- so code and package retrieval reaches roughly four times as many agent users as search does. 
Thus, the realistic injection surface is not a search result, but a README, issue body, docstring, or setup script that the agent retrieves and reads as part of its normal workflow. Therefore injections are difficult to distinguish from legitimate instructions. The surface extends further through agent extensions: 16.3\% of agent users run skills, plugins, MCP servers, or hooks, including 8.7\% with skills or plugins installed, 3.2\% running an MCP server, and 2.4\% executing their own hook scripts. These extensions introduce third-party code or instructions that execute with the user's facility identity, including access to allocations, group-writable project directories, and job-submission rights. They could waste compute, exfiltrate data, poison dependencies  and the shared environments.


\begin{table}[t]
\centering
\caption{Destructive or cleanup commands issued by users.}
\label{tab:destructive-commands}
\setlength{\tabcolsep}{3pt}
\begin{tabular}{@{}llrrr@{}}
\toprule
\textbf{Operation} & \textbf{User Type} & \textbf{Events} &
\textbf{Users} & \textbf{Events/User} \\
\midrule
\multirow{2}{*}{\texttt{rm -f}}
    & Agent & 23,655 & 235 & 100.7 \\
    & Human & 5,458  & 306 & 17.8 \\
\midrule
\multirow{2}{*}{Process kill}
    & Agent & 5,612 & 140 & 40.1 \\
    & Human & 24    & 8   & 3.0 \\
\midrule
\multirow{2}{*}{\shortstack[l]{\texttt{git reset --hard}\\
                                or \texttt{git clean -fd}}}
    & Agent & 1,726 & 209 & 8.3 \\
    & Human & 141   & 31  & 4.5 \\
\bottomrule
\end{tabular}
\vspace{-1em}
\end{table}


\noindent\textbf{Agents can expose services and leak credentials.} Agents nowadays run on login nodes, which are shared by many users and sensitive information can become easily visible to others. Process command lines may be inspected through \texttt{/proc} or tools such as \texttt{ps aux}, while shell history, job scripts, logs, and environment dumps often persist on shared filesystems. Agents can unintentionally place secrets into all of these places. For example, a bearer token passed to \texttt{curl} or a personal access token embedded in a \texttt{git clone} URL
can become visible from the command, shell history or job output afterward. Because agents also tend to print commands, inspect logs, and copy outputs into their context, the same secret can be propagated further into files, repositories, or model requests. \emph{In our measurement, 127 API keys are exposed in the duration of three weeks.}

Agents can also expose services unintentionally. When asked to ``serve the results'', an agent may launch Jupyter on \texttt{0.0.0.0}, making it reachable from other machines or users without realizing the security implications. None of these actions are obviously malicious, and they are often convenient shortcuts for completing a task. \emph{On a shared cluster, however, those shortcuts can become an exposed proxy of the cluster.}

\subsection{Opportunities: System-Level Guardrails, Provenance
Tracking, and Safe Secrets}

\noindent\textbf{Enforce guardrails for dangerous operations.} 
Safety checks should operate beneath the agent, at the system boundary, because agents tend to work around the restriction. A hook that intercepts shell actions before execution could reject credentials in \texttt{argv}, \texttt{chmod 777} on shared paths, sandbox widening, or compute launched on login nodes. 
However, command-level filtering alone is insufficient
because the same effect may be reached indirectly through a script, interpreter, or another system call. Kernel mechanisms such as Landlock~\cite{landlock} and seccomp illustrate why stronger isolation is typically enforced at the resource or system-call boundary rather than solely by inspecting commands. The open question is how to combine lightweight, agent-aware hooks with enforceable system-level isolation while avoiding false positives that block legitimate scientific workflows.

\noindent\textbf{Treat fetched content as untrusted and scope third-party extensions.} Content fetched from external sources should not be treated as equivalent to instructions from the user. The system could track the provenance of text entering the agent's context and propagate a form of \emph{taint} when externally sourced content influences later decisions. Privileged actions, such as changing permissions, could then require an untainted source of authority rather than being authorized solely by fetched text.

The same principle should apply to extensions such as skills, plugins, and MCP servers.
Extensions should have explicit identities and declared capabilities, and be admitted according to site policy rather than inheriting unrestricted access. The open challenge is how to preserve flexibility 
while preventing untrusted components from silently acquiring access to credentials, shared data, or cluster-wide resources.

\noindent\textbf{Make secret handling safe and limit service exposure.} Secrets can leak through many channels, including command output, logs, shell history, and model requests. A key challenge is whether sensitive values can be reliably detected and redacted across all of these paths: overly aggressive redaction can break legitimate workflows, while permissive filtering risks exposing credentials.

Agents often need network access to fetch data, making complete isolation impractical, while unrestricted binding to non-loopback interfaces can expose services. 
Facilities must decide whether to require explicit approval, block such bindings by default, or mediate them through site policy, and make sure agents follow that policy correctly.

\section{Conclusion}

We presented a large-scale measurement of agents on a production HPC
cluster. The measurements show that agents are
already a significant workload and interact with the facility differently from humans: faster, finer-grained, and more
iterative, pursuing open-ended goals through trial-and-error campaigns.
From these trends we distilled the challenges agents bring to four aspects of
the facility, namely compute, storage, agent memory, and safety. For each of the four aspects, we discussed opportunities and open research
questions. We believe the co-design of agents and HPC systems is necessary
and will benefit both the users and the cluster management.
 
\section{Acknowledgement}
We thank Paul Edmon from Harvard FAS Research Computing for his contributions to setting up our data collection infrastructure, and for sharing valuable feedback on our earlier ideas. ChatGPT is used to polish the language.

\bibliographystyle{IEEEtran}
\bibliography{refs}

\end{document}